\documentclass[twocolumn,secnumarabic,amssymb, nobibnotes, aps, prl, superscriptaddress, nobalancelastpage,longbibliography,nofootinbib]{revtex4-1}

\usepackage[utf8]{inputenc}

 \usepackage{graphicx}
\usepackage{bm}
\usepackage{amsmath} \usepackage{braket} \usepackage{epsfig}
\usepackage{tensor}
\usepackage{CJKutf8}
\usepackage[version=4]{mhchem}
\usepackage{titlesec}
\usepackage{ifthen}
\usepackage{sidecap}
\usepackage{listings}
\usepackage[para,online,flushleft]{threeparttablex}

\usepackage{hyperref}
\usepackage[all]{hypcap}
\usepackage{float}
\usepackage{multibib}

\makeatletter
\def\@bibdataout@aps{%
\immediate\write\@bibdataout{%
@CONTROL{%
apsrev41Control%
\longbibliography@sw{%
    ,author="08",editor="1",pages="1",title="0",year="1"%
    }{%
    ,author="08",editor="1",pages="1",title="",year="1"%
    }%
  }%
}%
\if@filesw \immediate \write \@auxout {\string \citation {apsrev41Control}}\fi
}
\makeatother

\usepackage{xcolor}
\definecolor{pastelgray}{rgb}{0.81, 0.81, 0.77}
\definecolor{beaublue}{rgb}{0.9, 0.9, 0.93}

\newcommand{\NNLO}{$\Delta$NNLO$_\text{GO}$}
\newcommand{\Fy}{Fy($\Delta r$, HFB)}
\newcommand{\remove}[1]{}

\begin{document}

\title{Universal trend of charge radii of even-even Ca-Zn nuclei}

\author{Markus Kortelainen}
\affiliation{Department of Physics, University of Jyväskylä, PB 35(YFL) FIN-40014 Jyväskylä, Finland.}

\author{Zhonghao Sun}
\affiliation{Department of Physics and Astronomy, University of Tennessee, Knoxville, Tennessee 37996, USA.}
\affiliation{Physics Division, Oak Ridge National Laboratory, Oak Ridge, TN 37831, USA.}

\author{Gaute Hagen}
\affiliation{Physics Division, Oak Ridge National Laboratory, Oak Ridge, TN 37831, USA.}
\affiliation{Department of Physics and Astronomy, University of Tennessee, Knoxville, Tennessee 37996, USA.}

\author{Witold Nazarewicz}
\affiliation{FRIB Laboratory and Department of Physics and Astronomy , Michigan State University, East Lansing, Michigan 48824, USA.}

\author{Thomas Papenbrock}
\affiliation{Department of Physics and Astronomy, University of Tennessee, Knoxville, Tennessee 37996, USA.}
\affiliation{Physics Division, Oak Ridge National Laboratory, Oak Ridge, TN 37831, USA.}

\author{Paul-Gerhard Reinhard}
\affiliation{Institut für Theoretische Physik, Universität Erlangen, Erlangen, Germany.}

\begin{abstract} 
Radii of nuclear charge distributions carry information about the strong and electromagnetic forces acting inside the atomic nucleus. While the global behavior of nuclear charge radii is governed by the bulk properties of nuclear matter, their local trends are affected by quantum motion of proton and neutron nuclear constituents. The  measured differential charge radii $\delta\langle r^2_c\rangle$ between neutron numbers $N=28$ and $N=40$ exhibit a universal pattern as a function of $n=N-28$ that is independent of the atomic number. Here we analyze this remarkable behavior in even-even nuclei from calcium to zinc using two state-of-the-art theories based on quantified nuclear interactions: the ab-initio coupled cluster theory and nuclear density functional theory. Both theories reproduce the smooth rise of differential charge radii and their weak dependence on the atomic number.
By considering a large set of isotopic chains, we show  that this trend can be captured by just two parameters: the slope and curvature of ${\delta\langle r^2_c\rangle(n)}$. We demonstrate that
these parameters show appreciable model dependence, and the statistical  analysis indicates that they are not correlated with any single model property, i.e., they are impacted by both bulk nuclear properties as well as shell structure.
\end{abstract} 

\maketitle


{\it Introduction.---} High-precision measurements of nuclear charge radii
offer unique information on the structure of atomic nuclei and fundamental symmetries of nature \cite{Counts2020,Berengut2020}. In particular, the precise data on   variations of charge radii with proton and neutron numbers   shed light on elusive aspects of nuclear behavior, such as superfluidity \cite{Hammen18,Gorges2019,Miller2019,deGroote2020}, shell structure \cite{Koszorus2021,reponen2021}, and correlations \cite{Yordanov2020,malbrunot2021}.

A case in point is  the calcium isotopic chain, in which the charge radii show an arch-like behavior with pronounced odd-even staggering between the neutron magic numbers $N=20$ and $N=28$, with the charge radius of $^{48}$Ca very close to the value in $^{40}$Ca \cite{Emrich1983}, followed by a steep rise resulting in
unexpectedly large charge radii of neutron-rich  isotopes \cite{Garcia2016}. Several structural effects contribute to this intricate pattern: smooth scaling of nuclear radii  with the nuclear mass number as $A^{1/3}$, configuration mixing~\cite{Caurier2001}, the zero-point motion associated with surface vibrations~\cite{Barranco1985}, nucleonic pairing \cite{ReinhardNazarewicz2017} in the presence of  particle continuum~\cite{Miller2019}, as well as  nucleonic charge form-factors and relativistic corrections~\cite{Friar1975}. In particular, the spin-orbit correction to nuclear charge density results in pronounced shell effects
attributed to the population of spin-unsaturated single-particle orbits
\cite{Friar1975,ReinhardNaz2021} that helps explaining the anomalous reduction of the charge radius in $^{48}$Ca.

The recent paper~\cite{Koszorus2021} showed that measured differential charge radii exhibit an element-independent steep increase beyond neutron number $N=28$ from potassium ($Z=19$) to iron ($Z=26$). 
The main  aim of the present study is to address this puzzling behavior  in even-even nuclei with atomic numbers $20\le Z\le 30$ and $N\le 40$. 
Experimentally, these nuclei exhibit a variety of structures, 
ranging from patterns characteristic of spherical nuclei in the vicinity of the nuclear magic numbers (here $N,Z=20$ or $N,Z=28$) to collective behavior attributed to well-deformed, open-shell  systems as, e.g., $^{60}$Fe. 
In these ranges of particle numbers, protons and neutrons move predominantly in the $1p0f$ shell-model orbitals, which, above $N=28$, form a single pseudo-SU(3) shell \cite{Arima1969,Raju1973}
whose orbits have very similar radial behavior \cite{Troltenier1994}.

From a phenomenological perspective, changes in charge radii are attributed to  core polarization effects due to valence nucleons~\cite{Bohr52} resulting in nuclear deformations \cite{Brix1958}. 
Excellent insights were obtained by  the seniority-model approach to nuclear radii~\cite{Zamick1971,Talmi1984},
which is expected to work particularly well in the upper $1p0f$  shell because of the aforementioned pseudo-SU(3) symmetry.
Moreover, because of the presence of nucleonic pairing, the variation of $\delta\langle r_c^2\rangle$ with particle numbers is expected to be smooth in even-even nuclei.

In the generalized seniority  picture, for a given semi-magic isotopic chain, the attractive  interaction between the core nucleons and valence neutrons leads to a simple parabolic pattern of differential radii in even-$N$ isotopes \cite{Zamick1971,Talmi1984}:
\begin{equation}\label{eq:arc}
\delta\langle r_c^2\rangle^{A_{\rm m},A_{\rm m}+n}=
an+bn^2,
\end{equation}
where $n$ is defined as the number of
neutrons above the shell closure at $N=N_{\rm m}$ in the magic nucleus $A_{\rm m}$, and $a$ and $b$ are constants.
Experimentally, such a regular behavior has been seen in a number of even-even isotopes in the  $0f_{7/2}$ region above $N_{\rm m}=20$ and is not limited to semi-magic systems~\cite{Avgoulea2011,Wohlfahrt1981}. 
For instance, for the calcium chain, assuming  identical charge radii  of $^{40}$Ca  and $^{48}$Ca, Eq.~(\ref{eq:arc}) yields an arch-like trend: 
\begin{equation}\label{eq:arcCa}
\delta\langle r_c^2\rangle^{40,40+n}=\frac{n(8-n)}{16} \delta\langle r_c^2\rangle^{40,44}.
\end{equation}
Considering the nuclei with $N>28$, the generalized seniority scheme predicts   that  differential radii should behave according to Eq.~(\ref{eq:arc}) with $N_{\rm m}=28$.


{\it Theoretical methods.---} To understand the observed trends, we carried out a theoretical analysis using  models that are capable of describing deformed nuclei. The first approach is the ab-initio coupled cluster (CC) theory~\cite{Hagen2014} that allows for systematically improvable calculations based on realistic Hamiltonians with nucleon–nucleon and three-nucleon potentials.  The second approach is based on nuclear density functional theory (DFT)~\cite{Bender2003}. 

The comparison between ab-initio and DFT results for the root-mean-square (rms) charge radii
$\sqrt{\langle r_c^2\rangle}$
and differential  charge radii
$\delta\langle r_c^2\rangle^{A,A'}
\equiv \langle r_c^2\rangle^{A'}-\langle r_c^2\rangle^{A}$, closely related to isotope shifts,  
has been presented in several recent papers \cite{Hagen2015,deGroote2020,Koszorus2021,malbrunot2021}. In general, the energy-functional-based approach provides a more accurate description of nuclear global properties, including the total charge radii, while the local variations, e.g., differential radii, are better captured by Hamiltonian-based  methods.


In our CC calculations, we employed the recently developed two- and three-nucleon {\NNLO} interaction~\cite{jiang2020} with a cutoff of 394~MeV. This interaction from chiral effective field theory (EFT) is based on pion exchange, short-ranged contacts and, in contrast to many potentials derived within the framework of chiral effective field theory~\cite{ RevModPhys.81.1773, MACHLEIDT20111}, the {\NNLO} interaction also includes the effects of $\Delta$-resonance  degrees of freedom~\cite{kaiser1998,krebs2007,epelbaum2008,piarulli2015}. The resulting potential yields an accurate description of bulk properties in finite nuclei and the saturation point and symmetry energy of nuclear matter~\cite{jiang2020}. 

We performed single-reference CC calculations in the singles and doubles (CCSD) approximation~\cite{kuemmel1978,bartlett2007}, using a natural-orbital basis computed from an axially symmetric, parity conserving, Hartree-Fock (HF) state~\cite{sam2020}, in a model space consisting of 15 harmonic oscillator shells ($N_{\rm max}=14$). This approach extends the previous computations of charge radii~\cite{Hagen2015} to open-shell nuclei and allows us to describe deformed nuclei.
Correlations beyond the mean field are captured by the ensuing CC computations. Our approach precludes the inclusion of effects from triaxial shapes (which are not expected to be significant in the nuclei we computed). We also note that the restoration of rotational invariance, i.e., the projection onto good total angular momentum, remains a challenge that we do not address in this work. The effects of the lacking symmetry restoration on the ground-state energy can be estimated from the projection of the HF energy~\cite{sam2020}. 

The HF density matrix is initialized by filling the last occupied single-particle orbits from low to high values of $|j_z|$, which  yields a prolate deformed nucleus. After the HF equations are solved we compute a more accurate density matrix using second-order many-body perturbation theory~\cite{PhysRevC.99.034321} and obtain the natural orbital basis by diagonalizing the density matrix. 
Following Ref.~\cite{hoppe2021} the normal-ordered Hamiltonian in the two-body approximation~\cite{hagen2007a,roth2012} is then  truncated to a smaller model space (consisting of as many single-particle states as a harmonic-oscillator space with $N_{\rm max}^{\rm nat}=12$) according to the occupation numbers of the natural orbitals with respect to the Fermi level. Table~\ref{natcon} shows the CCSD ground-state energy and the charge radius of $^{64}$Zn in model spaces with oscillator spacings of $\hbar\omega=12$ and $\hbar\omega=16$\,MeV.
We see that both the charge radius and the ground-state energy converge rapidly in the natural-orbital basis, and $N_{\rm max}^{\rm nat}=12$ is sufficient for a converged result within 1\,MeV. The results presented in this work are based on the oscillator spacing $\hbar\omega=16$\,MeV and $N_{\rm max}^{\rm nat}=12$. 

\begin{table}[ht]
\centering
\caption{Convergence of the CCSD ground-state energy ($E$) and charge radii ($R_{\mathrm{ch}}$) in the natural-orbital basis for the nucleus $^{64}$Zn. The full space has $N_{\rm max}=14$, and truncated model spaces have dimensions equal to oscillator spaces with $N_{\rm max}^{\rm nat}=6, 8, 10, 12$.}
\label{natcon}
\begin{tabular}{c|ll|ll}
\hline\hline
& \multicolumn{2}{c}{$\hbar\omega$=12~MeV}   
& \multicolumn{2}{|c}{$\hbar\omega$=16~MeV}     \\
$N_{\rm max}^{\rm nat}$ & $E$(MeV) & $R_\text{ch}$(fm)& $E$(MeV) & $R_\text{ch}$(fm) \\\hline
6  & $-473.731$  & 3.857 & $-474.445$  & 3.848   \\
8  & $-513.502$  & 3.882 & $-515.685$  & 3.869   \\
10 & $-520.787$  & 3.896 & $-523.355$  & 3.882 \\
12 & $-521.746$  & 3.900 & $-524.384$  & 3.886 \\                                                          \hline\hline  
\end{tabular}
\end{table}

\begin{figure}[htb]
\includegraphics[width=1.0\columnwidth]{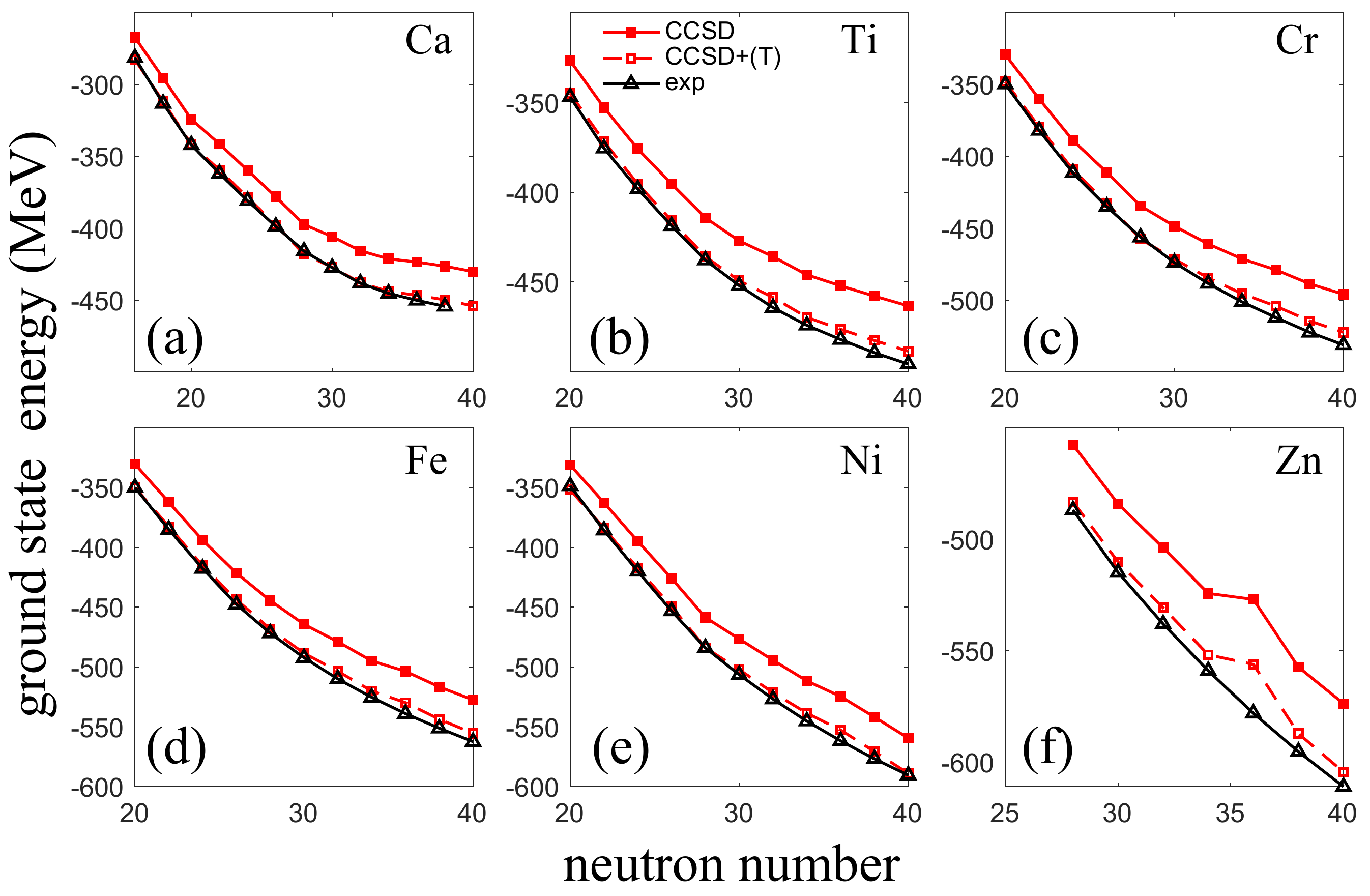}
\caption{ Ground-state energies of nuclei computed in this work using the CCSD approximation (a)--(f). Also shown are CCSD with estimated triples correction of 12\% of the correlation energy and compared to data. }
\label{gs-ergs}
\end{figure}

Figure~\ref{gs-ergs} shows the ground-state energies of nuclei computed in this work in the CCSD approximation. For natural orbitals, the estimated triples correction is about 12\% of the correlation energy. The nucleus $^{66}$Zn is an outlier and this might indicate that an axially-symmetric prolate deformed reference state is not most adequate here.

In the calculation of rms charge radii from the CC proton-point radii, we took
the  nucleonic  charge  radii $\langle r_p^2\rangle=0.709~\text{fm}^2$ and $\langle r_\text{n}^2\rangle=-0.106~\text{fm}^2$  \cite{filin2020}, and the Darwin-Foldy term
 $\langle r_\text{DF}^2\rangle=0.033$~fm$^2$.
We also included  the spin-orbit correction calculated within CC theory~\cite{Hagen2015}.
We have estimated that our ab-initio CC results for charge radii carry $\pm 2\%$ uncertainty due to model-space and cluster-operator truncations~\cite{sam2020}.


Our DFT calculations were carried out using two different energy density
functionals (EDFs): the Skyrme parametrization SV-min \cite{Kluepfel2009} and the Fayans functional parametrization {\Fy}
\cite{ReinhardNazarewicz2017,Miller2019}. Both functionals  were optimized to the same large set of experimental observables from Ref. \cite{Kluepfel2009}. In addition, the {\Fy} included  differential  radii of Ca isotopes. Its extended  pairing functional turned out to be essential for reproducing charge radii in the Ca isotopic chain~\cite{Miller2019} and the kinks in charge radii at magic numbers \cite{Gorges2019}.

The DFT calculations were done with the SkyAx \cite{Reinhard2021} and HFBTHO \cite{HFBTHO} solvers allowing for deformed solutions. For each nucleus, deformation energy minimum was located.
All calculations with HFBTHO used oscillator basis functions up to $N_{\rm sh}=20$. Pairing with SkyAx was done with a soft cutoff in single-particle
space with a Woods-Saxon profile \cite{Kri90a} reaching 15 MeV
above the Fermi level with a smoothing with of 1.5 MeV. Results for
spherical nuclei have been counter-checked with the spherical DFT solver, which was used for the calibration of the both
functionals. Because of different description of the continuum space, in HFBTHO pairing was renormalized to the results of 
spherical Fayans code. 

\begin{figure}[htb]
\includegraphics[width=1.0\columnwidth]{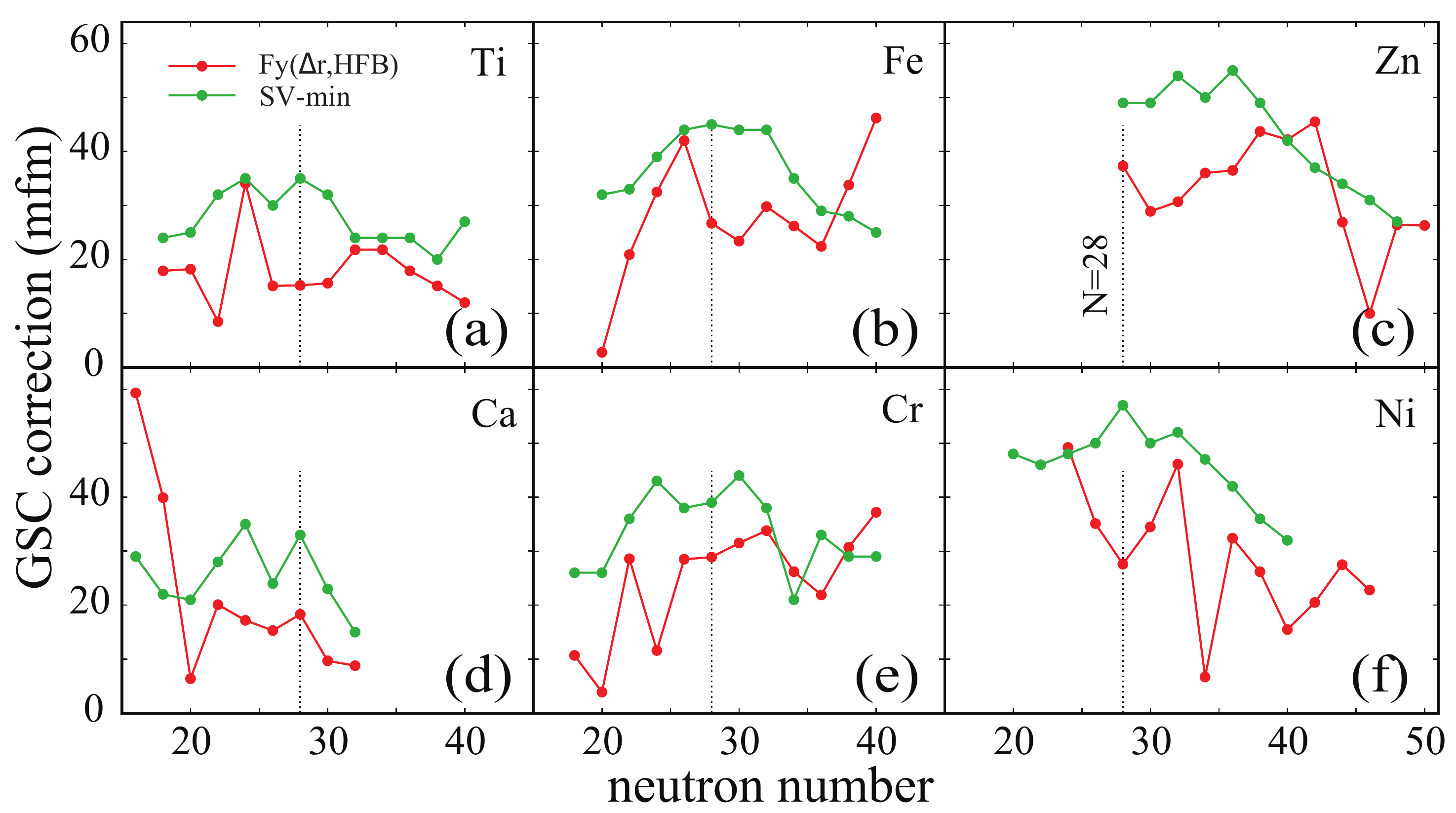}
\caption{Contribution to the charge radius from collective ground state correlations caused by low-lying quadrupole states for the two DFT parametrizations used  here.}\label{corrections}
\end{figure}

DFT parameterizations being calibrated to empirical data carry statistical uncertainties from the calibration strategy, coined statistical errors. They  can be estimated using the standard linear regression technique based on least squares  \cite{Dobaczewski2014}. 
Another source of error are systematic effects stemming from  limitations of the model. The largest systematic effects  can be traced back to correlations associated with  low lying collective modes, particularly the quadrupole vibrations that cannot be incorporated into a smooth density functional~\cite{Dreizler1990}. We have computed the effect of collective ground state correlations from low lying $2^+$ states on charge  radii~\cite{Kluepfel2008}. The results are shown in Fig.~\ref{corrections}. 
%
%
We take that as estimate for the systematic error. The DFT  error bands for charge radii shown in the following  represent  statistical and systematic uncertainties.

The charge radii from DFT calculations were obtained as in Ref.~\cite{ReinhardNaz2021}
by folding the point charge distribution with the intrinsic nucleon form factors. In this way, the contributions from nucleonic charge form-factors and relativistic corrections are automatically included.


\begin{figure}[!htb]
\includegraphics[width=1.0\linewidth]{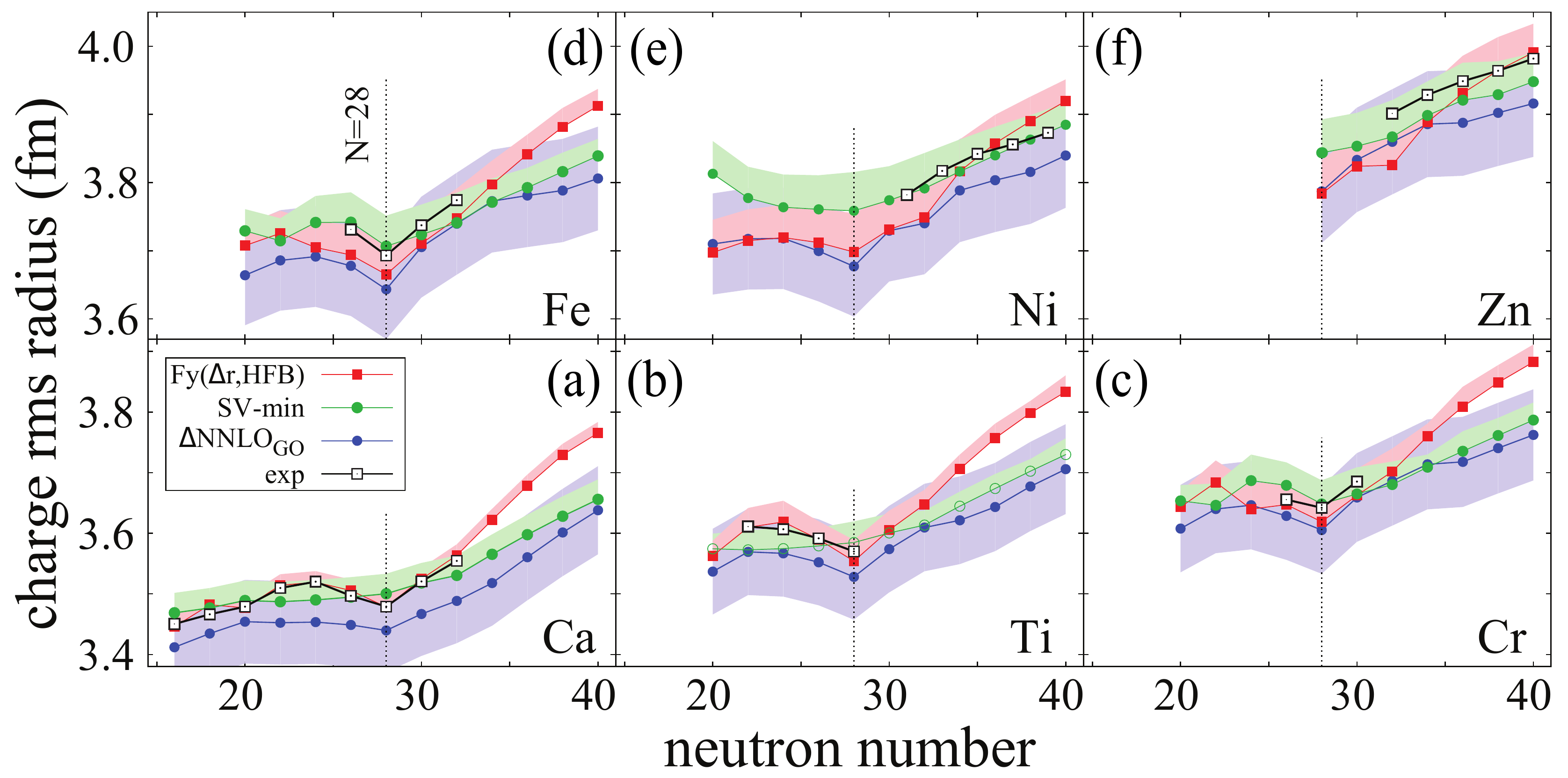}
\caption{ Charge rms radii along the isotopic chains of Ca (a), Ti (b), Cr (c), Fe (d), Ni (e), and Zn (f).
Results of CC and DFT calculations are compared to experimental data.
The error bars for the DFT calculations are shown as red and green shaded areas and the
error bands for the CC calculations are shown as blue shaded area.
Experimental data were taken from Refs.~\cite{Garcia2016,Miller2019} (Ca), \cite{Gangrsky2004} (Ti), \cite{Minamisono2016} (Fe), \cite{Kaufmann2020,malbrunot2021} (Ni),  \cite{Xie2019} (Zn), and Ref.\cite{Angeli2013}.
}\label{totalradii}
\end{figure}

{\it Results for charge radii.---} The results of our CC and DFT calculations for the rms charge radii for  even-even isotopic chains between calcium and zinc are shown in Fig.~\ref{totalradii}, including our estimates for theoretical uncertainties.
The Fayans functional {\Fy} provides a  detailed description of experimental data for the Ca, Ti, and Cr isotopic chains. This is not surprising as {\Fy} was optimized to the trends of charge radii in the selected  Ca isotopes. For the  light Ni, and Zn nuclei, {\Fy} systematically underestimates the charge radii. As discussed in Ref.~\cite{deGroote2020}, this is primarily related to the pairing in the $pf$-shell
region: the pairing Fayans functional adjusted using data from the calcium region is too strong
in heavier nuclei. The functional SV-min reproduces the charge radii data well, especially for the heavier isotopic chains. Its main deficiency is the underestimation of kinks in charge radii around magic gaps \cite{ReinhardNazarewicz2017,Gorges2019}. Generally, the employed {\NNLO} interaction tends to predict slightly reduced  radii compared to DFT results and experiment. 

No pronounced irregularities in charge radii  are seen at neutron numbers $N=32, 34$, which have been predicted to be magic by some models. This conclusion is consistent with the findings of Ref.~\cite{Koszorus2021}, which studied this effect in the potassium isotopes.

\begin{figure}[!htb]
\includegraphics[width=0.7\linewidth]{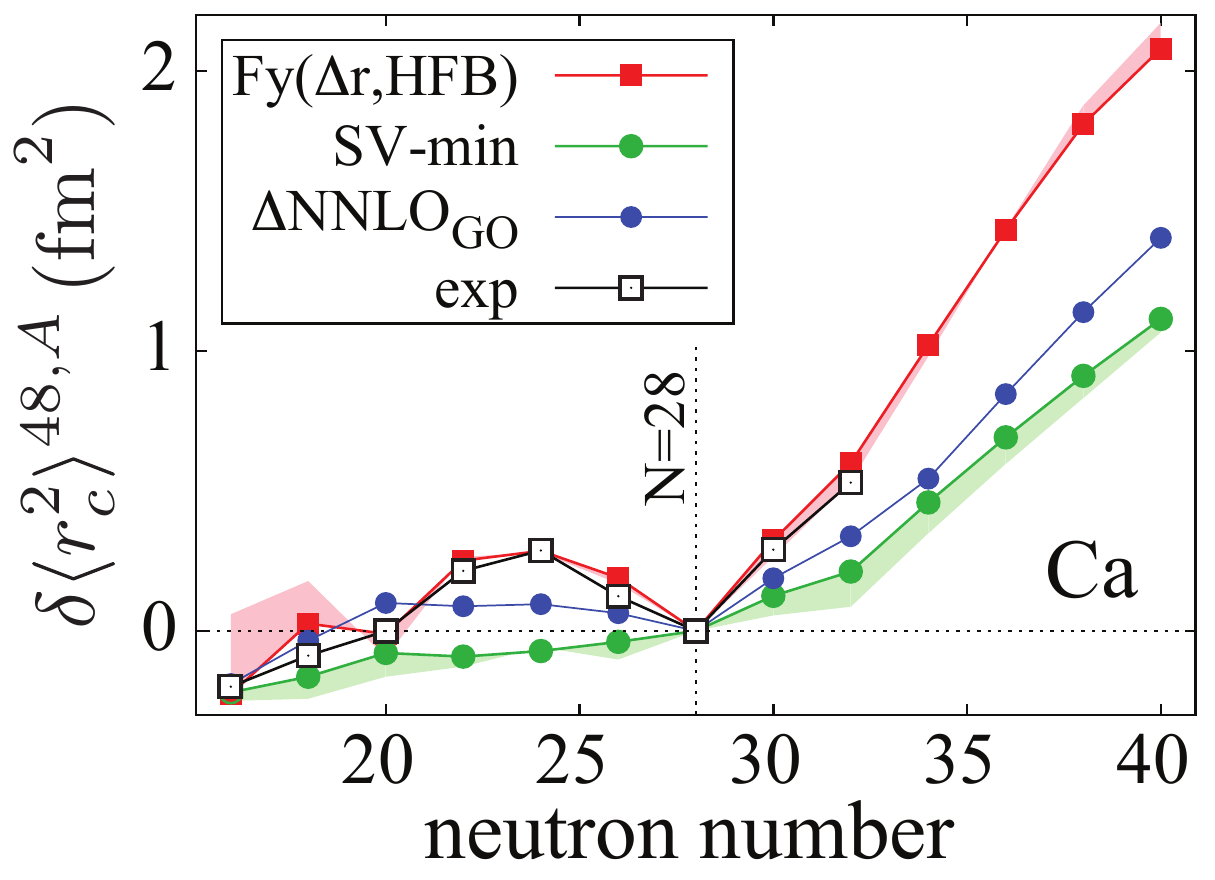}
\caption{ Differential radii of the Ca isotopes. Theoretical uncertainties for CC calculations have been omitted.}\label{isoshifts}
\end{figure}

As discussed earlier, the Ca chain represents the particular challenge for nuclear theory. This is illustrated in 
Fig.~\ref{isoshifts}, which compares our predictions for the  differential radii of the Ca isotopes with experiment. 
As expected, the local arch-like behavior  between $^{40}$Ca  and $^{48}$Ca, and a steep increase above $N=28$  are both well reproduced by {\Fy}, which has been constrained by experimental values 
$\delta\langle r^2_c\rangle^{40,48}$,  $\delta\langle r^2_c\rangle^{44,48}$, and  $\delta\langle r^2_c\rangle^{48,52}$, cf. also Eq.~(\ref{eq:arcCa}). The model SV-min yields a rather smooth monotonic increase of  $\delta\langle r^2_c\rangle^{48,A}$
with neutron number. The results of {\NNLO} fall in between.
As indicated by Eq.~(\ref{eq:arcCa}) the value of 
the charge radius of the neutron open-shell nucleus $^{44}$Ca 
 is important for understanding the trend in the Ca radii in the  $0f_{7/2}$ region. Here, we note that the properties of $^{44}$Ca are strongly impacted by neutron pairing and that the specific pairing interaction of {\Fy} has been crucial for describing the experimental trend.
All three models predict an increase of   the charge radius for $N>28$. The rate of this increase is, however, model dependent.

\begin{figure}[!htb]
\includegraphics[width=1.0\linewidth]{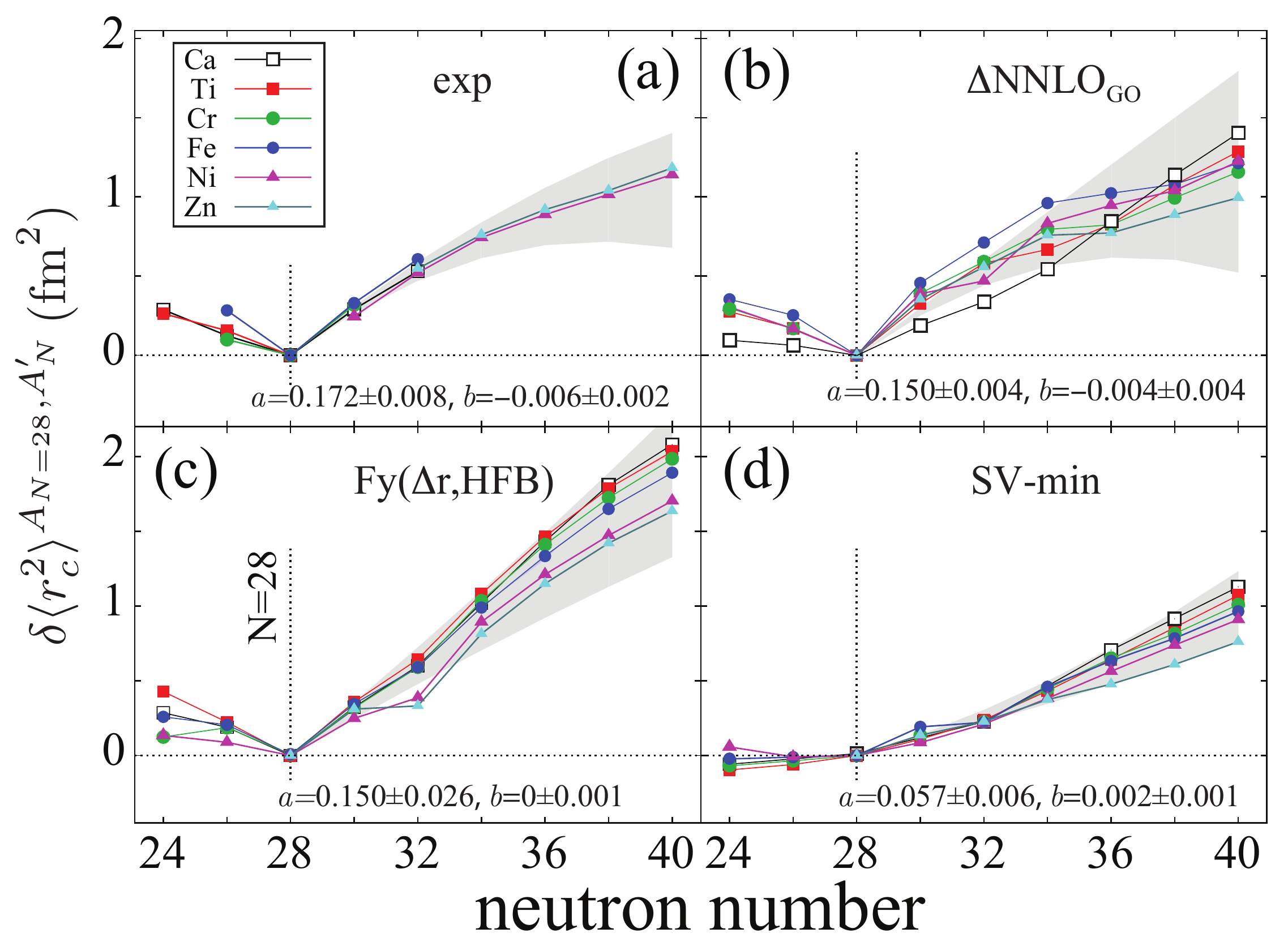}
\caption{ Experimental (a) and theoretical (b--d) differential radii for the even-even Ca-Zn isotopic chains relative to the value of $\langle r^2_c\rangle$  at $N=28$. Theoretical error bars have been omitted. The values of $a$ and $b$ coefficients of Eq.~(\ref{eq:arc}) (in fm$^2$) obtained to the quadratic fit to $Z$-averaged differential radii for $N>28$ are shown together with corresponding uncertainties (marked by a gray shading).}\label{modelradii}
\end{figure}

Our main result is presented in Fig.~\ref{modelradii}, which shows $\delta\langle r^2_c\rangle$ by comparing all isotopic chains in the same panel. As noted in Ref.~\cite{Koszorus2021},  the measured charge radii beyond
$N = 28$  exhibit a common increasing trend that is irrespective of the atomic number. This remarkable property is also present in our calculations: the predicted differential radii primarily depend on one quantity only:
the number $n=N-28$ of valence neutrons outside the $N=28$ gap. However, the  predicted values of $\delta\langle r^2_c\rangle$  exhibit model-dependent patterns 
as a function of $n$.

The degree of $Z$-independence seen in experimental data is quite astonishing. It would be very interesting to see whether the experimental extension of the current limits of charge radii data for neutron-rich Ti, Cr, and Fe  isotopes will confirm the pattern seen in Fig.~\ref{modelradii}. Due to the limited data supply at the experimental side and remaining errors at the theoretical side, the isotopic spread seen in theoretical results is greater than in experiment. Still, considering the scale of the deviations, the degree of isotopic consistency is quite good:
the predicted model trends in $\delta\langle r^2_c\rangle$ 
represent excellent benchmarks for theory.
A similar  $Z$-independent pattern of $\delta\langle r^2_c\rangle$ above $N=28$ has recently been obtained in the  Green’s function approach~\cite{Soma2021} for $Z=20,22,24$ and $N\le 36$ with the interaction NNLO$_{\rm sat}$. Below we compare their results to ours.

In order to quantify the $n$-dependence, we fitted the $Z$-averaged values of $\delta\langle r^2_c\rangle$ to the parabolic expression (\ref{eq:arc}). 
The resulting $a$ and $b$ coefficients are shown in  Fig.~\ref{modelradii} for experimental data and for  each model. One sees that the ($a,b$) values 
describing experimental differential charge radii are very close to those obtained in {\NNLO}: in both cases the pattern of charge radii is described by a  parabola which is concave down. This result confirms the earlier observation that the local trends in charge radii in the $pf$ nuclei are well described by ab-initio theory with chiral EFT interactions~\cite{deGroote2020,Koszorus2021}. We note  that our analysis of the   Green’s function results from Ref.~\cite{Soma2021},  yields $a=0.084\pm 0.006$\,fm$^2$, and $b=0.003\pm 0.001$\,fm$^2$, i.e.,
their slope parameter is well below experimental   and CC+{\NNLO} values.

For {\Fy}, the average curvature coefficient $b$ is very close to zero.
For SV-min, on the other hand $b$ is positive, i.e., the resulting pattern is concave up. To see whether the parameters $(a,b)$ contain information about specific model properties,  we carried out a statistical  least-square regression study following the methodology described in Ref.~\cite{Schuetrumpf2017}. This analysis requires the control of all model parameters simultaneously.

The EDF parameters, characterizing its bulk properties,
can be conveniently expressed through properties of symmetric nuclear matter; those are: the equilibrium density
$\rho_{0}$, the energy-per-nucleon at equilibrium $E/A$,
the incompressibility $K$, the effective mass $m^*/m$ characterizing the
dynamical isoscalar response, the symmetry energy $J$ and its slope $L$,  and the 
Thomas-Reiche-Kuhn sum-rule enhancement
$\kappa$ characterizing the dynamical isovector response, see
Ref.~\cite{Kluepfel2009} for definitions. In addition,
we consider two parameters characterizing surface properties ($b_{\rm surf}$ and $b'_{\rm surf}$), two parameters 
characterizing spin-orbit terms ($b_{\rm ls}$ and $b'_{\rm ls}$), and three  pairing parameters ($V_{\rm pair,pr}$, $V_{\rm pair,ne}$,  and $\rho_{\rm pair}$).

\begin{figure}[htb]
\includegraphics[width=1.0\columnwidth]{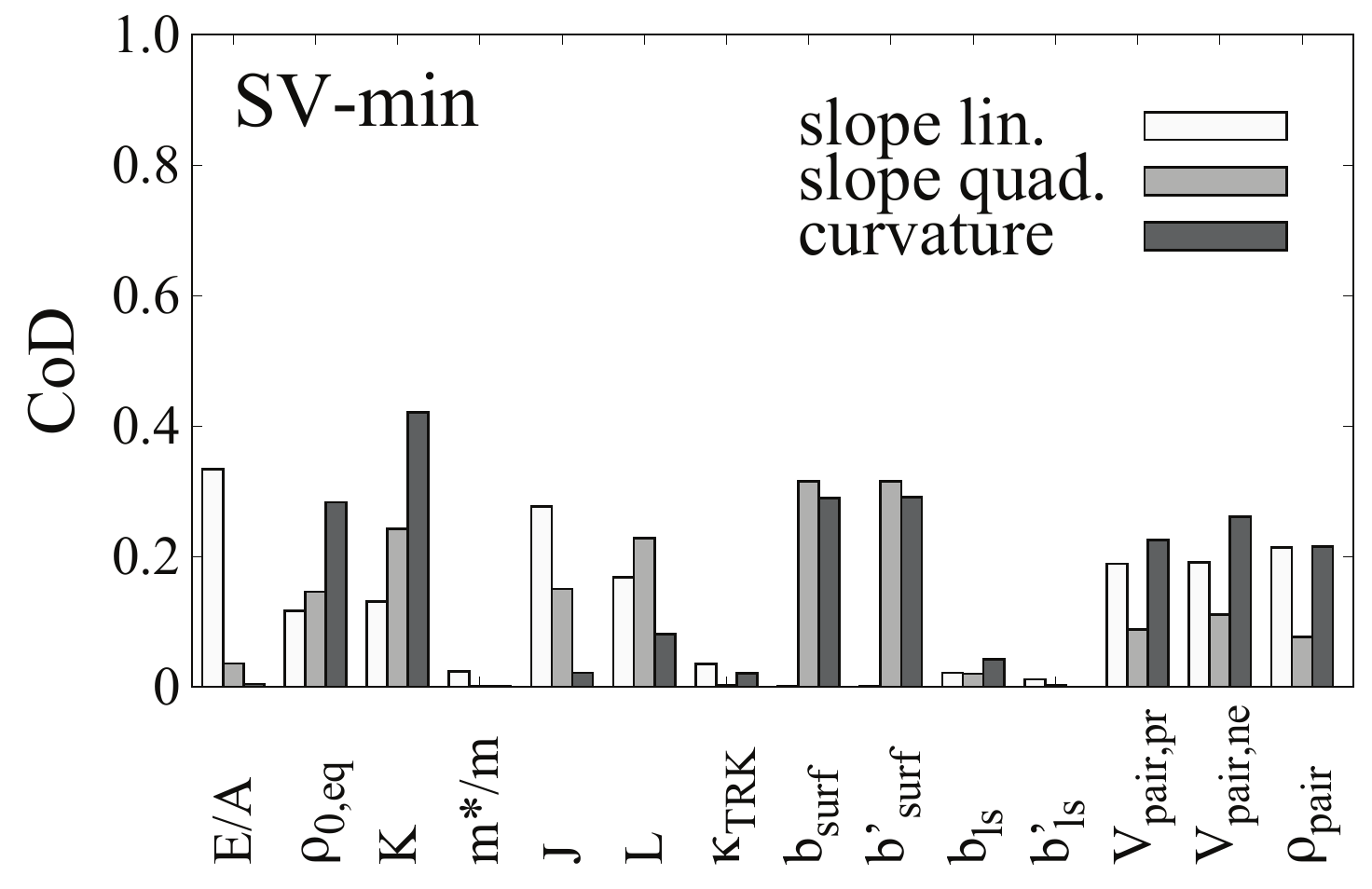}
\caption{Coefficients of determination between the parameters of SV-min and the slope and curvature coefficients of equation~(\ref{eq:arc})  fitted to the $Z$-averaged values of $\delta\langle r^2_c\rangle$ of  Fig.~\ref{modelradii} with $28\le N\le 40$. }
\label{correlations}
\end{figure}

Figure~\ref{correlations} displays the results of our statistical analysis in 
terms of the coefficients of determination between the slope and curvature coefficients of equation~(\ref{eq:arc}) 
fitted to the $Z$-averaged values of $\delta\langle r^2_c\rangle$ of Fig.~\ref{modelradii} with $N\ge 28$. The coefficients of determination is
a square of the bivariate correlation coefficient; it contains information on how well one quantity is determined by another one. 
 
We carried out two fits. In the first variant (slope lin.), we estimated the average slope assuming no curvature. In the second variant, we estimated both the slope $a$ (slope quad.) and curvature  $b$. Our analysis 
indicates that $(a,b)$ values  are not correlated with one single model parameter, i.e., they are ``distributed observables'' and they are impacted by both bulk nuclear properties and shell structure.


{\it Discussion.---} In conclusion, the regular pattern of  the measured differential radii of even-even Ca-Zn nuclei beyond neutron number $N = 28$ has been analyzed by means of CC and DFT calculations extended to the open-shell, deformed nuclei. While the absolute charge radii are more accurately described by nuclear DFT, the local trends are very well modelled by CC calculations. Both theories reproduce the smooth rise of charge radii and their weak dependence on the atomic number. We have demonstrated that experimental and theoretical charge radii for these nuclei can be well parameterized in terms of a parabolic function of the  number $n$  of valence neutrons outside the $N=28$ magic gap. In particular, CC calculations reproduce both the slope and curvature parameters describing the experimental $\delta\langle r^2_c\rangle(n)$ dependence. To trace back the values of these parameters to properties of nuclear forces remains a theoretical challenge for future investigations.

{\it Acknowledgements.}
Discussions with Á. Koszorús and G. Neyens are acknowledged. We thank V. Som{\`a} for providing us with Gorkov-NNLO$_{\rm sat}$ results for charge radii. This material is based upon work supported by the U.S.\ Department of Energy, Office of Science, Office of Nuclear Physics under award numbers DE-SC0013365, DE-FG02-96ER40963, and  DE-SC0018083, DE-SC0018223 (NUCLEI SciDAC-4 collaboration). 
This work has been partially supported by the Academy of Finland under the Academy Project No. 339243.
Computer time was provided by the Innovative and Novel Computational Impact on Theory and Experiment (INCITE) programme. This research used resources of the Oak Ridge Leadership Computing Facility located at Oak Ridge National Laboratory, which is supported by the Office of Science of the Department of Energy under contract No. DE-AC05-00OR22725.
We acknowledge the CSC-IT Center for Science Ltd. (Finland) for the 
allocation of computational resources.

\end{document}